\documentstyle[aps,prl,twocolumn,graphicx]{revtex}
\newcommand{\be}{\begin{eqnarray}}
\newcommand{\ee}{\end{eqnarray}}

\begin{document}
\draft

\twocolumn[\hsize\textwidth\columnwidth\hsize\csname @twocolumnfalse\endcsname
\title{High Temperature Charge-Ordering Fluctuation in Manganites}
\author{K. H. Kim$^{1,*}$, M. Uehara$^{1,\dagger}$, S-W. Cheong$^{1,2}$}
\address{$^1$Department of Physics and Astronomy, Rutgers University, Piscataway, New Jersey 08854}
\address{$^2$Bell Laboratories, Lucent Technologies, Murray Hill, New Jersey 07974}
\date{submitted to PRL at Nov. 30,1999}
\maketitle
\begin{abstract}
High temperature-({\it T}) studies of La$_{1-x}$Ca$_{x}$MnO$_{3}$ show that charge
and phonon transport is significantly suppressed in a narrow doping region
of {\it x}$\approx $1/2 and in a wide {\it T} range up to 900 K, which accompanies
anomalous broadening of x-ray Bragg peaks. Contrarily, ferromagnetic (FM)
correlation above Curie {\it T} is sharply enhanced at {\it x}$\approx $3/8. 
All these observations indicate the presence of short-range charge ordering (CO)
correlation at high {\it T}, possibly in the form of a FM ``zigzag", a small
segment of the CE-type CO state. The decoupling and coupling of the FM
zigzags for {\it x}$\approx $3/8 and 1/2, respectively, is consistent with our
results. 
\end{abstract}
\pacs{PACS numbers: 75.30.Vn, 72.20.Pa, 72.15.Eb, 72.80.Tm}
\vskip0.5pc]

\newpage

Recent extensive studies have shown that the formation of static or dynamic
charge/spin stripes is the generic feature of the doped Mott insulators \cite
{1}. The static charge ordering (CO) with stripe correlation has been
observed in layered nickelates \cite{2}, perovskite manganites \cite{3} and
doped orthoferrites \cite{4}. In addition, the presence of the static
correlation of the charge/spin stripes has been observed in layered cuprates
in the region where superconductivity is suppressed \cite{5}. Furthermore,
it has been proposed that the incommensurate inelastic neutron peaks
observed in superconducting cuprates are due to the dynamic charge/spin
stripe correlation \cite{6}. Therefore, understanding the relationship
between the static or dynamic charge/spin stripe correlation with other
physical properties such as superconductivity in doped Mott insulators is
one of the challenging issues in current condensed matter physics. 

In mixed-valent manganites, orbital degree of freedom associated with Mn$^{%
\text{3+}}$ ions, in addition to charge and spin degrees of freedom, plays
an important role. The static charge/orbital ordering with stripe patterns
is now well established, especially in La$_{\text{1-}{\it x}}$Ca$_{{\it x}}$%
MnO$_{\text{3}}$ with {\it x }$\geq $ 0.5 at low temperature ({\it T}) region. CO
in manganites occurs as periodic arrays of the sheet-like arrangement of Mn$%
^{\text{3+}}$ ions \cite{7}. In this scheme, the CO state of La$_{\text{0.5}}
$Ca$_{\text{0.5}}$MnO$_{\text{3}}$ is special in the sense that the density
of Mn$^{\text{3+}}$-Mn$^{\text{4+}}$ pairs is highest. In La$_{\text{0.5}}$Ca%
$_{\text{0.5}}$MnO$_{\text{3}}$, all of the charge, orbital and spin degrees
of freedom freeze into the so-called CE-type stable configuration below 180
K (for heating) \cite{8,9}.

Even though static, long-range striped ordering of charge, orbital, and spin
degrees of freedom in manganites is well established, the dynamic or
short-range correlation of these degrees of freedom is poorly understood. In
the simple double exchange (DE) mechanism, hopping of $e_{\text{g}}$
electron of Mn$^{\text{3+}}$ induces ferromagnetic (FM) coupling between
localized $t_{\text{2}g}$ spins \cite{10}. Thus, the FM correlation in La$_{%
\text{1-}{\it x}}$Ca$_{{\it x}}$MnO$_{\text{3}}$ is supposed to vary as {\it %
x}(1-{\it x}), and should be optimized at {\it x}=0.5, basically because of
the maximum number of neighboring Mn$^{\text{3+}}$-Mn$^{\text{4+}}$ pairs.
However, the Curie {\it T} ({\it T}$_{\text{C}}$) peaks mysteriously at {\it x}$\approx $%
3/8 \cite{9}. The DE-type FM correlation is, naturally, suppressed by charge
localization, which is a prerequisite for CO. Furthermore,
superexchange-type magnetic coupling between localized spins sensitively
depends on the local orbital configuration, and the orbital degree of
freedom can be coupled with the lattice through the Jahn-Teller (JT)
mechanism. Thus, the spatial or temporal fluctuations of charge, spin,
orbital, and lattice degrees of freedom ought to be strongly coupled to each
other.

In order to reveal the thermal fluctuation nature of these degrees of
freedom, we have carefully studied the various physical properties of La$_{%
\text{1-}{\it x}}$Ca$_{{\it x}}$MnO$_{\text{3}}$ with {\it x} near 1/2 and
3/8 at high {\it T} regions. Surprisingly, the suppression of electronic
conductivity and phononic thermal conductivity exists in a narrow doping
range of {\it x}$\approx $1/2, which persists up to 900 K, far above static
ordering {\it T}. On the other hand, we found that FM correlation at room {\it T}, above
{\it T}$_{\text{C}}$, is sharply optimized at {\it x}=3/8. All of these behaviors
can be understood in terms of varying degree of CO correlation with Ca
doping level.

High quality polycrystalline specimens of La$_{\text{1-}{\it x}}$Ca$_{{\it x}%
}$MnO$_{\text{3}}$ with various {\it x} near 1/2 and 3/8 have been prepared
with the standard solid state reaction with an identical synthesis
condition. Resistivity ($\rho $) was measured with the standard four probe
method with accurate geometry from 4 to 900 K, and absolute thermal
conductivity ($\kappa $) was measured from 8 to 310 K with the steady state
method (employing a radiation shield). Magnetization (M) was measured by
using Quantum Design SQUID magnetometer, and x-ray powder diffraction data
were taken by using Rigaku diffractometer. M data show that the system with
0.45$\leq ${\it x}$\leq $0.50 undergoes FM transition at 240-220 K,
accompanied by antiferromagnetic (AFM) CO at 160-180 K (for heating) \cite
{9,11}. The specimens with {\it x} near 3/8 undergo only FM transition at $%
\sim $270 K, and CO transition around 200 and 220 K was observed in the {\it %
x}=0.52 and 0.55 samples, respectively.

Figure 1(a) shows the $\rho $({\it T}) curves for La$_{\text{1-}{\it x}}$Ca$_{{\it %
x}}$MnO$_{\text{3}}$ with {\it x} near 0.5 from 4 to 900 K. First, we
discuss the behaviors of $\rho $ with the variation of {\it x} at low {\it T}. For 
{\it x}=0.48, the FM metallic phase is dominant below 220 K even though a
short-range CO phase probably coexists as suggested by a broad hump and
small hysteresis at 100-180 K \cite{12}. However, with {\it x} approaching
0.5 from below, the CO state stabilizes at low {\it T}. Thus, $\rho $ near {\it x}%
=0.5 at low {\it T} increases systematically with {\it x}, and shows the
insulating {\it T}-dependence when {\it x}$\geq $0.485. As shown in Fig. 2(a)
(open circles), $\rho $(100 K) shows such a systematic increase with {\it x}
approaching 0.5 from below, consistent with a crossover from the FM metallic
to the CO insulating states. Once CO is stabilized at low {\it T} region for {\it x%
}$\geq $0.5, $\rho $(100 K) becomes insensitive on {\it x}.

It is intriguing to note that as evident in Fig. 1(a), $\rho $ of La$_{\text{%
0.5}}$Ca$_{\text{0.5}}$MnO$_{\text{3}}$ at {\it T}%
%TCIMACRO{\TEXTsymbol{>}}%
%BeginExpansion
\mbox{$>$}%
%EndExpansion
{\it T}$_{\text{CO}}$ is considerably larger than that of any neighboring
compositions, and this behavior persists up to 900 K. To unveil this
unexpected behavior in detail, we have systematically changed the chemical
doping level with fine spacing of {\it x} near 0.5. The $\rho $(300 K) vs. 
{\it x} plot in Fig. 2(a) summarizes the results, showing a clear maximum at 
{\it x}=0.5. Moreover, the $\rho $(900 K) vs. {\it x} plot confirms that the
maximum behavior at {\it x}=0.5 persists up to, at least, 900 K. A previous
study revealed that an adiabatic small polaron model, with $\rho $=$\rho _{_{%
\text{0}}}${\it T}exp($E_{\text{a}}$/{\it k}$_{\text{B}}${\it T}), describes high {\it T} $%
\rho $ of La$_{\text{1-}{\it x}}$Ca$_{{\it x}}$MnO$_{\text{3}}$ in broad
doping and {\it T} ranges (0$\leq ${\it x}$\leq $1 and $\sim $300K$\leq ${\it T}$%
\leq $1100K) \cite{13}. Here, $E_{\text{a}}$ represents the activation
energy of small polarons, i.e. the potential barrier that polarons must
overcome to hop to the next site. The inset of Fig. 1 shows that ln($\rho $%
/{\it T}) vs. 1/{\it T} plot of our data at high {\it T} region is quite linear, corroborating
with the adiabatic small polaron model. Interestingly,  $E_{\text{a}}$ is
systematically enhanced at {\it x}=0.5, as shown in Fig. 2(b). Therefore,
the strong charge localization tendency at {\it x}=0.5 up to very high {\it T},
far above {\it T}$_{\text{CO}}$, is closely associated with the enhancement of
polaron activation energy.

To gain further insights into understanding this surprising result, we have
measured {\it T}-dependent $\kappa $ (Fig. 1(b): heating). At {\it T}s above {\it T}$_{\text{C}%
}$, $\kappa $({\it T}) for {\it x} 
%TCIMACRO{\TEXTsymbol{<} }%
%BeginExpansion
\mbox{$<$}%
%EndExpansion
0.5 increases linearly with increasing {\it T}, which is unusual in crystalline
solids. First, we point out that the electronic $\kappa $ estimated from $%
\rho $ by using the Wiedemann-Franz law is negligible. It has bee known that
the electronic $\kappa $ of adiabatic small polarons is also very small \cite
{14}. In addition, an earlier work on FM manganites has shown that the
anomalous behavior of $\kappa $ , i.e., a linear increase above {\it T}$_{\text{C}%
} $ is well explained by the phononic $\kappa $, coupled with large
anharmonic lattice distortions \cite{15}. Thus, it is evident that at high {\it T}
region of La$_{\text{1-}{\it x}}$Ca$_{{\it x}}$MnO$_{\text{3}}$ ({\it x}$%
\approx $0.5), the electronic contribution to $\kappa $ is negligible, and
the phonon contribution, possibly related to large anharmonic lattice
distortions, dominates the measured $\kappa $.

The abrupt $\kappa $ increase near {\it T}$_{\text{C}}$ in $\kappa $({\it T}) of {\it x}%
=0.485, 0.49, and 0.50 of Fig. 1(b) is due to the reduced lattice
distortions in the FM-metallic state, and this $\kappa $ increase at {\it T}$_{%
\text{C}}$ becomes smaller with {\it x} approaching 0.5. For {\it x}$\leq $%
0.5, $\kappa $ tends to decrease at {\it T}$_{\text{CO}}$, which can be attributed
to the large (JT-type) lattice distortion associated with CO. For {\it x} 
%TCIMACRO{\TEXTsymbol{>} }%
%BeginExpansion
\mbox{$>$}%
%EndExpansion
0.5, $\kappa $ shows only slight slope changes near {\it T}$_{\text{CO}}$, as seen
in the $\kappa $ data of {\it x}=0.52 and 0.55. As shown in Fig. 2(c), $%
\kappa $ at low {\it T} region decreases systematically with {\it x} approaching
0.5 (due to the stabilization of the CO state), and remains small when {\it x%
}%
%TCIMACRO{\TEXTsymbol{>}}%
%BeginExpansion
\mbox{$>$}%
%EndExpansion
0.5. We note that even if the $\kappa $ values of {\it x}=0.52 and 0.55 are
similar to that of {\it x}=0.5 at low {\it T}, they become considerably larger
than that of {\it x}=0.5 for {\it T} 
%TCIMACRO{\TEXTsymbol{>} }%
%BeginExpansion
\mbox{$>$}%
%EndExpansion
{\it T}$_{\text{CO}}$. This behavior is well illustrated in the $\kappa $(300 K)
vs. {\it x} plot (solid circles in Fig. 2(c)), demonstrating a clear minimum
at {\it x}=0.5. This suppression of $\kappa $(300 K) at {\it x}=0.5
correlates well with the $\rho $ peaking near 0.5 at 300 K. Therefore, our
results in Figs. 1 and 2 show that the lattice thermal conductivity as well
as charge transport is suppressed in the high {\it T} region of the half-doped
manganite.

Directly related with the suppression of phononic $\kappa $, there exists a
slight, but noticeable broadening of the x-ray Bragg peaks for {\it x}=0.5
at room {\it T}. One example of the broadened x-ray peaks is shown in Fig. 3,
displaying the compositional change of the (242) Bragg peak (in the
orthorhombic P{\it bnm} notation) of x-ray powder diffraction at 300 K. The
(242) Bragg peak, centered at 2$\theta \approx $69.3 $^{\circ }$ for {\it x}%
=0.5, changes its position to higher angles as {\it x} increases. The left
and right sides of the (242) Bragg peak are due to the (004)-(400) peaks and
K$_{\alpha \text{2}}$ of the (242) and (004)-(400) peaks, respectively. As
evident in Fig. 3, the peak width, $\Gamma $, of the central (242) peak is
considerably broad at {\it x}=0.5. To extract the {\it x}-dependence of $%
\Gamma $, we have fitted the intensity profiles as a sum of three Gaussian
peaks (by neglecting the weak K$_{\alpha \text{2}}$ peaks of (004) and
(400)). The solid squares and triangles in Fig. 2(d) represent fitting
results for $\Gamma $ and the center position of the (242) peak,
respectively. The center position increases almost linearly with {\it x},
indicating the linear lattice contraction with increasing {\it x}. On the
other hand, $\Gamma $ shows a clear maximum at {\it x}=0.5. This broadening
of $\Gamma $ indicates a slight distribution of lattice constants in the
half-doped manganite at room {\it T} \cite{16}.

An early study of synchrotron x-ray scattering for La$_{\text{0.5}}$Ca$_{%
\text{0.5}}$MnO$_{\text{3}}$ showed a drastic broadening of all the Bragg
peaks in the FM region between {\it T}$_{\text{C}}$ and {\it T}$_{\text{N}}$ ($\approx ${\it T}%
$_{\text{CO}}$) \cite{16}. After this discovery, an electron diffraction
study showed that the fine scale ($\sim $100 \r{A}) coexistence of CO and FM
phases is responsible for the drastic broadening of x-ray Bragg peaks \cite
{3}. What our new data indicates is that this Bragg peak broadening for {\it %
x}$\approx $0.5 persists even at room {\it T}, far above {\it T}$_{\text{C}}$ and {\it T}$_{%
\text{CO}}$. Therefore, the Bragg peak broadening at room {\it T} suggests that
short-range CO exists in the paramagnetic state of the half-doped manganite.
We cannot rule out the possibility of dynamic correlation of CO at room {\it T}.
The spatial variation of lattice constants, indicated by the Bragg peak
broadening, can shorten phonon lifetime, and thus suppresses the phononic $%
\kappa $. The enhancement of $\rho $ up to 900 K in La$_{\text{0.5}}$Ca$_{%
\text{0.5}}$MnO$_{\text{3}}$ indicates that the spatial or dynamic CO
fluctuation probably persists even at {\it T} ranges much higher than long-range
ordering {\it T} \cite{17}. We emphasize that the observed Bragg peak broadening
indicates that various anomalous behaviors of {\it x}$\approx $0.5 are bulk
effects. In other words, the $\rho $ enhancement and the $\kappa $
suppression at {\it x}$\approx $0.5 are not due to, for example, grain
boundaries in our polycrystalline specimens. We also note that our findings
are not consistent with La/Ca ionic ordering because the La/Ca ordering
should reduce $\rho $.

Since CO correlation can influence magnetic correlation, we have measured
the evolution of magnetic susceptibility ($\chi \equiv $M/H) at room {\it T}
(above {\it T}$_{\text{C}}$) as a function of {\it x}. We found a few surprising
results as shown in Fig. 4(a). First, we found that $\chi $({\it T}) above {\it T}$_{%
\text{C}}$ roughly follows the Curie-Weiss law, and the Curie-Weiss {\it T} is FM
for all {\it x} studied. Consistent with earlier results, the {\it T}$_{\text{C}}$
vs. {\it x} plot shows a broad bump near {\it x}=3/8 \cite{9}. On the other
hand, $\chi $({\it x}) at 300 K sharply peaks at {\it x}=3/8. This
pronounced peaking of $\chi $({\it x}) above {\it T}$_{\text{C}}$ at the
commensurate carrier concentration of 3/8 suggests an extraordinary
possibility; the presence of short-range or dynamic correlation of
charge/orbital ordering in such a way as to produce special FM coupling in
addition to DE-type FM coupling.

It is important to understand how such short-range charge correlation can
promote FM coupling, which is seemingly counter-intuitive. In manganites, it
has been well established that the CE-type CO is very stable in broad {\it x}
ranges, at least at low {\it T}. For example, the CE-type CO has been commonly
observed in La$_{\text{1-}{\it x}}$Ca$_{{\it x}}$MnO$_{\text{3}}$ and Nd$_{%
\text{1-}x}$Sr$_{x}$MnO$_{\text{3}}$ for {\it x}$\approx $0.5. In addition,
the CE-type CO has been reported even when {\it x} deviates significantly
from 0.5. For example, the CE-type CO occurs in Pr$_{\text{1-}{\it x}}$Ca$_{%
{\it x}}$MnO$_{\text{3}}$ with 0.3%
%TCIMACRO{\TEXTsymbol{<}}%
%BeginExpansion
\mbox{$<$}%
%EndExpansion
{\it x}$\leq $0.5, and also in (La,Pr)$_{\text{5/8}}$Ca$_{\text{3/8}}$MnO$_{%
\text{3}}$ at low {\it T} \cite{12}. Thus, it is appealing to assume that
short-range or dynamic CO at high {\it T} of La$_{\text{1-}{\it x}}$Ca$_{{\it x}}$%
MnO$_{\text{3}}$ (0.2%
%TCIMACRO{\TEXTsymbol{<}}%
%BeginExpansion
\mbox{$<$}%
%EndExpansion
{\it x}$\leq $0.5) is also the CE-type. In the CE-type CO, there exist FM
zigzag chains (Fig. 4(b)), which couple to each other antiferromagnetically 
\cite{8,18,19}. It is conceivable that at high {\it T}, the CO correlation is so
short that the short-range CO state may contain only one short FM zigzag (Mn$%
^{\text{3+}}$- Mn$^{\text{4+}}$- Mn$^{\text{3+}}$- Mn$^{\text{4+}}$- Mn$^{%
\text{3+}}$: shown with dark hue in Fig 4(b)) with $\sim $11 \r{A}\ in
length or one short FM ``zig or zag'' (Mn$^{\text{3+}}$-Mn$^{\text{4+}}$-Mn$%
^{\text{3+}}$) with $\sim $5.5 \r{A}\ in length. Then, these extended
objects can enhance FM correlation overall. The short FM zigzag can be
considered as correlated polarons or a ferromagnetic polaron cluster, and
may exhibit dynamic nature. Note that the carrier concentration of the short
FM zigzag (zig or zag) corresponds to {\it x}=0.4 (1/3), which is close to 
{\it x}$\approx $3/8 for the enhanced FM correlation. However, if the range
of CO becomes slightly longer, then the zigzag may couple with the
neighboring zigzags antiferromagnetically so that FM correlation can be
reduced. This effect can be significant at {\it x}$\approx $0.5 where CO
tendency is strong because carrier concentration matches the CE-type
ordering \cite{20}. This remarkable scenario remains to be confirmed by
local probe measurements such as x-ray or neutron scattering experiments.

In conclusion, we discussed the suppression of electronic conductivity as
well as phononic thermal conductivity, and the broadening of Bragg peaks in
a narrow composition range near {\it x}=0.5, but in a very broad {\it T} range up
to 900 K. On the other hand, FM correlation is strongly enhanced for {\it x}
near 3/8 at high {\it T}. All these findings can be consistent with spatial or
temporal fluctuation of CO at high {\it T}. We have proposed an appealing scenario
of the presence of FM zigzags which can be coupled or decoupled, depending
on {\it x}. The ``decoupled'' short FM zigzag can enhance the overall FM
correlation at {\it x} near 3/8, and the AFM coupling of FM zigzags can
progressively increase with {\it x}, and be maximized at {\it x}$\approx $%
0.5 where charge localization tendency is strong. Our work clearly provides
a new paradigm for studying the fluctuation nature of CO correlation in
various doped Mott insulators, including mixed-valent manganites and
superconducting cuprates.

We are partially supported by the NSF-DMR-9802513. KHK. and MU are partially
supported by the KOSEF and by the JPSJ Fellowship, respectively.

Note: After submitting this paper, we became aware that the neutron
scattering study of La$_{\text{0.7}}$Ca$_{\text{0.3}}$MnO$_{\text{3}}$ by
Adams {\it et al.} \cite{21} are consistent with our proposed scenario of CO
correlation above {\it T}$_{\text{C}}$. They report the presence of the CE-type CO
correlation (correlation length of $\sim $10 \r{A}) above {\it T}$_{\text{C}}$.

\newpage 
\begin{figure}[tbp]
\caption{(a) {\it T}-dependent $\protect\rho $ curves (for heating and cooling) of
La$_{\text{1-}{\it x}}$Ca$_{{\it x}}$MnO$_{\text{3}}$ near {\it x}=0.5 from
4 to 900 K. The inset shows ln($\protect\rho $/{\it T}) vs. 1/{\it T} curves. (b) $%
\protect\kappa $({\it T}) (for heating) for La$_{\text{1-}{\it x}}$Ca$_{{\it x}}$%
MnO$_{\text{3}}$, {\it x}$\approx $0.5. The solid and dotted arrows show {\it T}$_{%
\text{C}}$ and {\it T}$_{\text{CO}}$ (determined from M({\it T}) and $\protect\rho $%
({\it T})), respectively. }
\label{Ref}
\end{figure}

\begin{figure}[tbp]
\caption{(a) and (c) $\protect\rho $ and $\protect\kappa $ values at 100 and
300 K for La$_{\text{1-}{\it x}}$Ca$_{{\it x}}$MnO$_{\text{3}}$ near $x$%
=0.5, respectively. (b) The activation energy of small polarons, {\it E}$_{%
\text{a}}$/{\it k}$_{\text{B}}$, vs. {\it x} plot, obtained from the inset
of Fig. 1. (d) The $x$-dependence of the peak width, $\Gamma $, and the
center position of the (242) x-ray Bragg peak at 300 K, estimated from
Gaussian fitting of the data in Fig. 3. The dotted and solid lines are
guides for the eyes.}
\label{Sig}
\end{figure}

\begin{figure}[tbp]
\caption{X-ray intensity profiles of the (242) Bragg peak (center) at 300 K
for La$_{\text{1-}{\it x}}$Ca$_{{\it x}}$MnO$_{\text{3}}$ near $x$=0.5. }
\label{TOPeak}
\end{figure}

\begin{figure}[tbp]
\caption{The {\it x}-dependence of $\protect\chi $ (closed squares) at 300 K
(left axis). The open circles show the {\it T}$_{\text{C}}$ variation determined
from $\protect\chi $({\it T}) (right axis). (b) A schematic of FM zigzag chains,
coupled antiferromagnetically each other. Open circles are Mn$^{\text{4+}}$
and the lobes show the $e_{\text{g}}$ orbitals of Mn$^{\text{3+}}$. ~$\sim $%
11 \AA\ FM zigzag is shown with dark hue. }
\label{XRD}
\end{figure}
\newpage
\clearpage
\vspace*{0cm}
\begin{center}
\hspace*{0cm}
\includegraphics[height=22cm,width=17cm,angle=0]{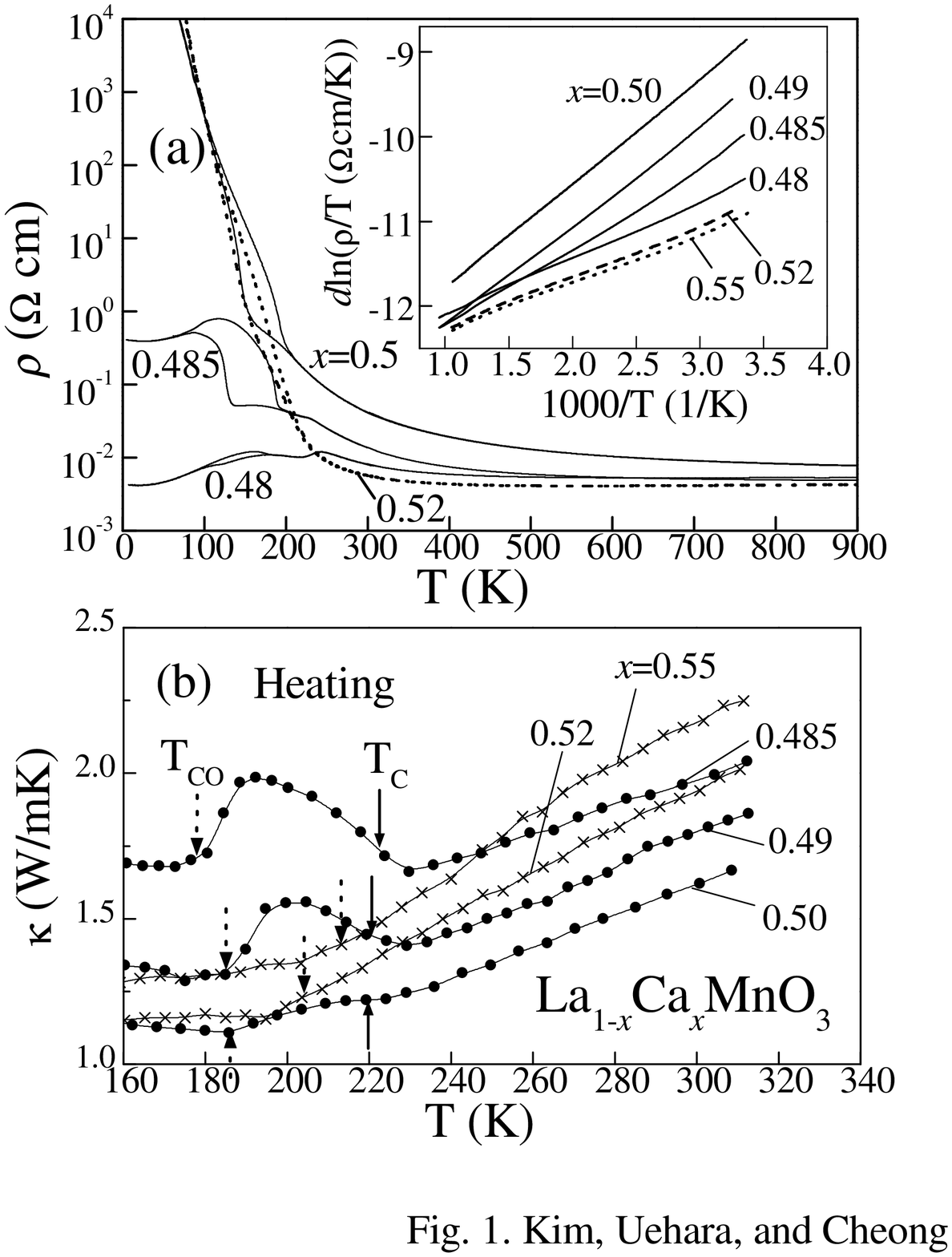}
%%\hspace*{1cm} \epsfig{file=mn327_fig4.eps,width=17truecm}
\end{center}
%% \vspace*{3.0cm}
% \large
%% \hspace{1cm} FIG. 1. K. H. Kim {\it et al.} 

\newpage
 \clearpage
\vspace*{0cm}
\begin{center}
\hspace*{0cm}
\includegraphics[height=22cm,width=17cm,angle=0]{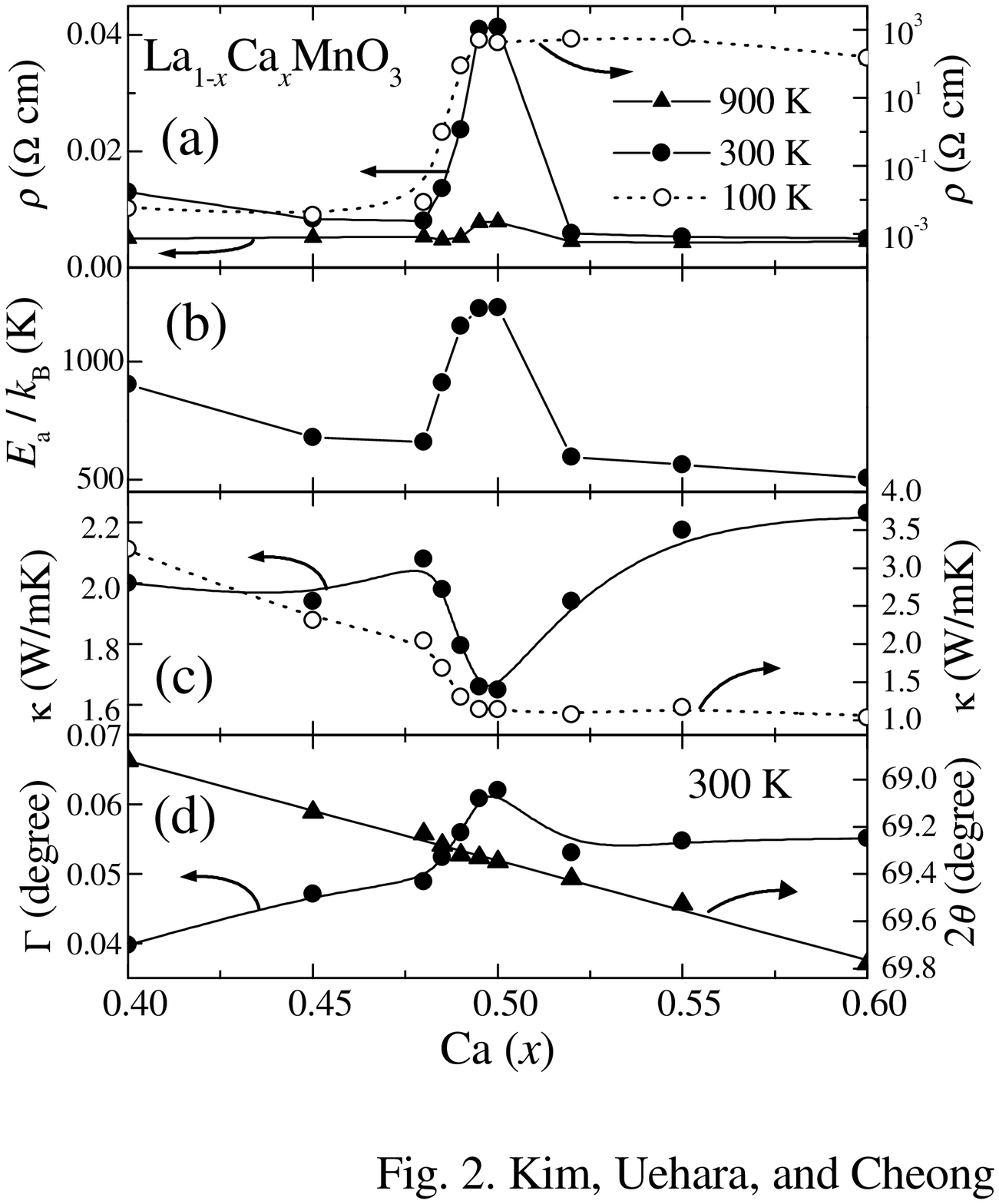}
%%\hspace*{1cm} \epsfig{file=mn327_fig4.eps,width=17truecm}
\end{center}
%% \vspace*{3.0cm}
% \large
%% \hspace{1cm} FIG. 2.  K. H. Kim {\it et al.}

\newpage
 \clearpage
\vspace*{0cm}
\begin{center}
\hspace*{0cm}
\includegraphics[height=18cm,width=22cm,angle=-90]{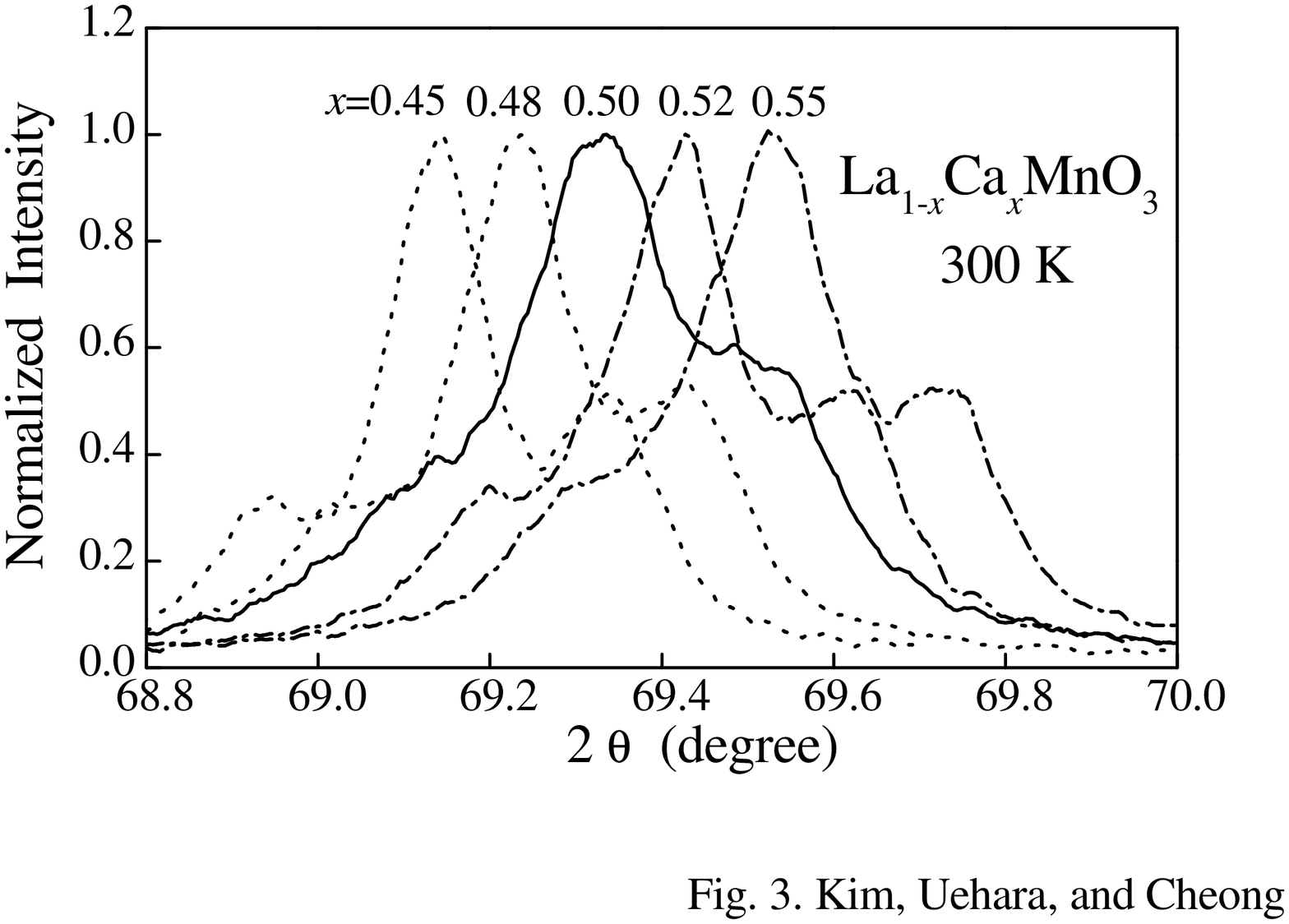}
%%\hspace*{1cm} \epsfig{file=mn327_fig4.eps,width=17truecm}
\end{center}
\vspace*{0.0cm}
% \large
%% \hspace{1cm} FIG. 3.  K. H. Kim {\it et al.}

\newpage
 \clearpage
\vspace*{0cm}
\begin{center}
\hspace*{0cm}
\includegraphics[height=22cm,width=17cm,angle=0]{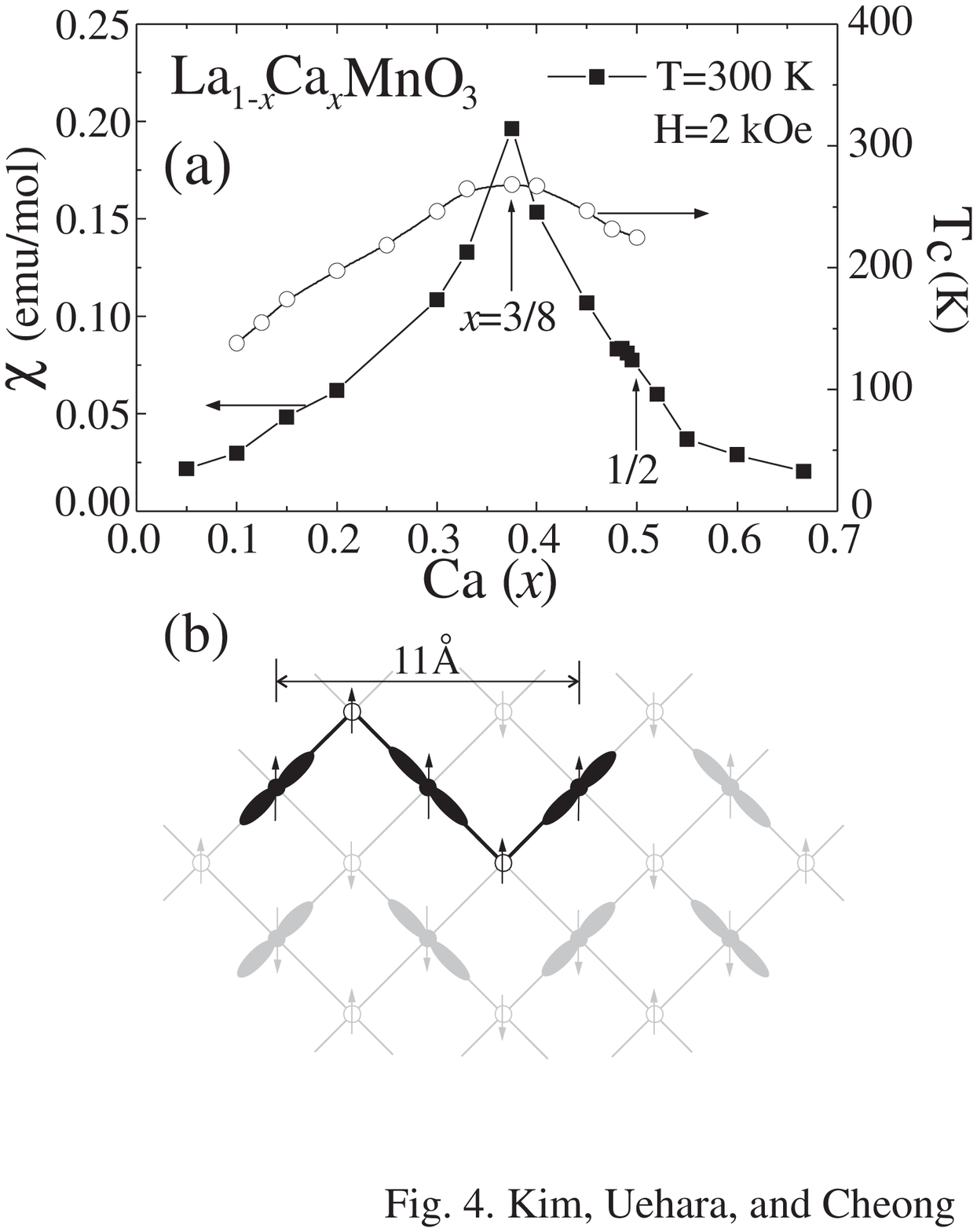}
%%\hspace*{1cm} \epsfig{file=mn327_fig4.eps,width=17truecm}
\end{center}
\vspace*{0.0cm}
% \large
%% \hspace{1cm} FIG. 4.  K. H. Kim {\it et al.}


\begin{references}
\bibitem{20}  We found that, in general, $\rho $(300 K) for {\it x}%
%TCIMACRO{\TEXTsymbol{<}}%
%BeginExpansion
\mbox{$<$}%
%EndExpansion
0.5 (including {\it x}$\approx $3/8 where {\it T}$_{\text{C}}$ is maximized) is
larger than that for {\it x}%
%TCIMACRO{\TEXTsymbol{>}}%
%BeginExpansion
\mbox{$>$}%
%EndExpansion
0.5 (including {\it x}$\approx $5/8 where {\it T}$_{\text{CO}}$ is optimized) even
though the ground state is metallic (insulating) for {\it x}%
%TCIMACRO{\TEXTsymbol{<}}%
%BeginExpansion
\mbox{$<$}%
%EndExpansion
(%
%TCIMACRO{\TEXTsymbol{>}}%
%BeginExpansion
\mbox{$>$}%
%EndExpansion
)0.5, which corroborates with short-range CO at high {\it T} for {\it x} near or
smaller than 0.5.
\end{references}
\end{document}